\documentstyle[12pt]{article}

\newcommand{\pl}{\partial} 
\newcommand{\be}{\begin{equation}} 
\newcommand{\ee}{\end{equation}} 
\newcommand{\bea}{\begin{eqnarray}} 
\newcommand{\eea}{\end{eqnarray}} 
\newcommand{\nn}{\nonumber} 
\newcommand{\p}[1]{(\ref{#1})} 
  
\begin{document}
\begin{flushright}
{\it To the memory of V.I. Ogievetsky}\\
\vskip.3cm
hep-th/9609090 \\
September 1996 
\end{flushright}
\centerline{\large\bf Harmonic superspace: new  
directions}
\vskip0.6cm
\centerline{{\large E.A. Ivanov}}
\vskip.3cm
\centerline{\it Bogoliubov Laboratory of Theoretical Physics, JINR,}
\centerline{\it 141 980, Dubna, Russian Federation}
\vskip.3cm
\centerline{\it Plenary Talk at 10th International Conference on} 
\centerline{\it PROBLEMS OF QUANTUM FIELD THEORY,}
\centerline{\it Alushta (Crimea, Ukraine) 13-18 May 1996} 
\vskip.5cm

\begin{abstract}{\small
\noindent We sketch recent applications of the harmonic 
superspace approach for off-shell formulations of $(4,4)$, $2D$ 
sigma models with torsion and 
for constructing super KdV hierarchies associated with "small" 
and "large" $N=4$ superconformal algebras. }
\end{abstract}
\vskip.5cm

\noindent{\bf 1. Introduction.} Harmonic superspace (HSS) has been 
proposed in 1984 by our Dubna group 
headed by late Viktor Isaakovich Ogievetsky \cite{GIKOS} as an efficient 
tool to treat theories with extended SUSY. 
This concept allowed to solve the long-standing problem of constructing  
off-shell superfield formulations of all $N=2$, $4D$ supersymmetric 
theories: $N=2$ matter (sigma models), $N=2$ super Yang-Mills and 
supergravity theories [1 - 4] as well as of 
$N=3$ super Yang-Mills 
theory \cite{GIKOSn3}. Later on, the same method was applied to purely 
bosonic problems to achieve a new formulation of the Ward construction 
for self-dual Yang-Mills fields \cite{GIOSap1} and to find unconstrained 
potentials for hyper-K\"ahler and quaternionic geometries 
\cite{{GIOSap},{quat}}. 

The essence of the harmonic (super)space approach consists in 
extending the original (super)manifold by some extra variables which 
parametrize the automorphism group of the given (super)manifold. In the 
$N=2$, $4D$ case it was the $SU(2)$ automorphism group 
acting on the 
Grassmann coordinates, the relevant additional variables being isospinor 
$SU(2)$ harmonics. The basic advantage of considering 
such extended 
manifolds is the possibility to single out in them a submanifold of 
lower dimension, the so-called ``analytic subspace''. In most examples 
the unconstrained functions on this subspace, analytic (super)fields, 
turn out to be the fundamental entities of the given theory. 

Since its invention, the HSS approach has been  
advanced and developed along several directions. 
One of new trends was applications 
to covariant quantization of superparticles and superstrings, 
as well as to constructing variants of the twistor-like formulation 
of these theories (see, e.g, refs. [9 - 13]). Another line 
was the further exploration of the relationships between complex 
target geometries and types of extended worldsheet supersymmetries 
in $2D$ sigma models. Using the 
$SU(2)$ harmonic (super)space language the most general action for 
sigma models with heterotic $(0,4)$ worldsheet supersymmetry has been 
found and the relevant bosonic target geometry has been revealed and 
studied \cite{{DKS},{DVal}} (in general, such sigma models 
possess a non-trivial 
torsion). Closely related development was the 
recent application of the same harmonic formalism for giving 
the $(0,4)$ superspace version of Witten's sigma model construction 
for ADHM instantons \cite{GS}. 
To set up general off-shell actions of torsionful 
$(4,4)$ supersymmetric sigma models, a new type of HSS, 
$SU(2)\times SU(2)$ one, has been recently proposed \cite{{IS},{Iv}}. 
One of the aims of this report is to give a brief account of the 
$SU(2)\times SU(2)$ harmonic approach and its applications in 
sigma models. 

One more interesting and perspective domain where the HSS techniques 
proved to be helpful is supersymmetric integrable 
models. Recently, the HSS methods were successfully 
used to find general solutions to self-dual super Yang-Mills and 
supergravity equations \cite{DO}, as well as to construct invariant 
actions for these systems \cite{Sok2}. This method also plays 
a central role in interpreting $N=3$ super Yang-Mills theory as an 
integrable $4D$ theory \cite{DO2} (the Lorentz $sl(2,C)$ harmonics are 
relevant in this case). An example of Lorentz invariant $2D$ 
integrable system with $(4,4)$ supersymmetry, the $(4,4)$ $SU(2)$ 
Liouville-WZNW theory \cite{{IK},{IKL}}, admits a nice description in 
$SU(2)\times SU(2)$ HSS \cite{IS}. Quite recently, 
the one-dimensional version of $SU(2)$ HSS was used for constructing 
and studying an $N=4$ superextension of the text-book example of 
integrable equation, the KdV one, based on the "small" 
$N=4$ superconformal algebra as the second hamiltonian 
structure \cite{{DI},{DIK}}. The survey 
of the harmonic superspace formulation of this algebra and 
associated $N=4$ KdV hierarchy is another subject of the present 
contribution. We will also briefly comment on the "large" 
$N=4$ SCA (with $SO(4)\times U(1)$ affine subalgebra) and the associate 
KdV hierarchy. In this case, the adequate formalism proves to be 
a version of the $SU(2)\times SU(2)$ HSS one. 

\vspace{0.3cm}
\noindent{\bf 2. SU(2)xSU(2) harmonic superspace}.
The $SU(2)\times SU(2)$ HSS 
is an extension of the 
standard real $(4,4)$ $2D$ superspace by two independent sets 
of harmonic variables  $u^{\pm 1\;i}$ and $v^{\pm 1\;a}$ 
($u^{1\;i}u^{-1}_{i} =
v^{1\;a}v^{-1}_{a} = 1$) associated with 
the automorphism groups $SU(2)_L$ and $SU(2)_R$ of the left and 
right sectors of $(4,4)$ supersymmetry \cite{IS} (see also 
\cite{BLR}). The corresponding 
analytic subspace is spanned by the following set of coordinates 
\be  \label{anal2harm}
(\zeta, u,v) = 
(\;x^{++}, x^{--}, \theta^{1,0\;\underline{i}}, 
\theta^{0,1\;\underline{a}}, u^{\pm1\;i}, v^{\pm1\;a}\;)\;,  
\ee
where we omitted the light-cone indices of odd coordinates (the first 
and second $\theta$s in \p{anal2harm} carry, respectively, the 
indices $+$ and $-$). The superscript ``$n,m$'' stands for two 
independent strictly preserved harmonic $U(1)$ 
charges, left ($n$) and right ($m$) ones.  
The additional doublet indices, $\underline{i}$ and $\underline{a}$, 
refer to two extra automorphism groups $SU(2)_L^{'}$ and $SU(2)_R^{'}$. 
Together with $SU(2)_L$ and $SU(2)_R$ they form the full $(4,4)$ 
supersymmetry automorphism group $SO(4)_L \times SO(4)_R$. 
We point out that the 
$SU(2)\times SU(2)$ harmonic superspace admits a manifest 
realization of the whole automorphism 
group of $(4,4)$ $2D$ supersymmetry. This is one of the reasons 
why a more general type of $(4,4)$ sigma models, 
those with torsion on the bosonic manifold, can be described within its 
framework. This type of sigma models is interesting mainly because 
they can provide non-trivial backgrounds for superstrings (see, e.g., 
ref. \cite{Luest}). 

More precisely, such a description becomes possible largely due 
to the fact 
that the $SU(2)\times SU(2)$ HSS provides a natural 
framework for off-shell formulation of the twisted $(4,4)$ multiplet. 
Until now, the latter was the basic object used for constructing 
sigma models of this type (actually, a subclass of them with mutually 
commuting left and right quaternionic structures \cite{{GHR},{RSS}}). 
It is described by a real analytic $SU(2)\times SU(2)$ harmonic 
superfield $q^{1,1}(\zeta,u,v)$ subjected to the harmonic constraints 
\cite{IS}
\be  \label{qconstr}
D^{2,0} q^{1,1} = D^{0,2} q^{1,1} = 0\;.
\ee
Here 
\bea
D^{2,0} &=& \partial^{2,0} + i\theta^{1,0\;\underline{i}}
\theta^{1,0}_{\underline{i}}\partial_{++}\;, 
\;\;D^{0,2} \;=\;\partial^{0,2} + i\theta^{0,1\;\underline{a}}
\theta^{0,1}_{\underline{a}}\partial_{--}  
\label{harm2der} \\
(\partial^{2,0} &=& u^{1 \;i}\frac{\partial} {\partial u^{-1 \;i}}\;,
\;\; 
\partial^{0,2} \;=\; v^{1 \;a}\frac{\partial} {\partial v^{-1 \;a}}) 
\nonumber 
\eea
are the left and right mutually commuting analyticity-preserving 
harmonic derivatives. These constraints leave in $q^{1,1}$ $8+8$ 
independent components that is just the irreducible off-shell 
component content of $(4,4)$ twisted multiplet. 

The general off-shell 
action of $n$ superfields $q^{1,1\;M}$  $(M=1,2,... n)$ is 
given by the following integral over the analytic superspace 
(\ref{anal2harm}) \cite{IS} 
\be
S_{q} = \int \mu^{-2,-2}\; h^{2,2}(q^{1,1\;M},u^{\pm1},v^{\pm1})\;,  
\label{qact}
\ee
$\mu^{-2,-2}$ being the relevant integration measure. 
The analytic superfield lagrangian $h^{2,2}$ is an arbitrary function 
of its arguments (the only restriction on its dependence on the 
harmonics $u$ and $v$ is the consistency with the external 
$U(1)$ charges $2,2$).  

Let us shortly characterize the relevant target bosonic geometry. 
$4n$ physical bosons $q^{ia\;M}_0(x)$ appear as the first component 
in the $u,v$ harmonic expansion of 
$q^{1,1\;M}= q^{ia\;M}_0(x)u^1_i v^1_a + ...$. 
The component physical bosons part of the action \p{qact} contains both 
the metric and torsion terms. Both the target metric and torsion are 
expressed in terms of the single symmetric $n\times n$ matrix function 
\bea
G_{M\;N} (q_0) = \int du dv\; g_{M\;N}(q^{1,1}_0, u, v)\;, \;\;\;
g_{M\;N}(q^{1,1}_0, u, v) = 
\frac{\partial^2 h^{2,2}}{\partial q^{1,1\;M} 
\partial q^{1,1\;N}}|_{\theta = 0}\;,
\label{auxfun} 
\eea
where $q_0^{1,1} \equiv q^{1,1} |_{\theta = 0}$. By its definition, 
this "metric" is a solution of the constraints 
\bea 
(a) \;\; \partial_{[M \;ia}G_{N]L} = 0,\;\;\;\; (b) \;\;
\partial_{M}^{\;\;ia}
\partial_{N\;ia} 
G_{LK} = 0\;. 
\eea
For four-dimensional targets there remains only one component 
in $G_{MN}$, so the relevant target metric is reduced 
to a conformal factor. The first condition in the above set is obeyed 
identically, while the second 
one becomes the Laplace's equation. This agrees with general conditions 
on the bosonic target in torsionful $(4,4)$ sigma models with 
four-dimensional targets \cite{{GHR},{CHS}}.

As a non-trivial example of the $q^{1,1}$ action with 
four-dimensional bosonic manifold we quote  
the action of $(4,4)$ extension of the $SU(2)\times U(1)$ 
WZNW sigma model  
\be  
S_{wznw} = \frac{1}{\kappa ^2} \int \mu^{-2,-2} \;
\left(\hat{q}^{1,1}\right)^2 \left(\frac{\mbox{ln}(1+X)}{X^2} 
- \frac{1}{(1+X)X} \right)\;. 
\label{confact} 
\ee 
Here  
\be 
\hat{q}^{1,1} = q^{1,1} - c^{1,1}\;,\;X = c^{-1,-1}\hat{q}^{1,1}\;,\; 
c^{\pm 1,\pm 1} = c^{ia}u^{\pm1}_iv^{\pm1}_a \;,\; 
c^{ia}c_{ia} = 2\;. 
\ee
Despite the presence of an extra quartet constant $c^{ia}$ in the 
analytic superfield lagrangian, the action (\ref{confact}) 
actually does not depend 
on $c^{ia}$ because it is invariant under arbitrary rigid
rescalings and $SU(2)\times SU(2)$ rotations of this constant. 
Its physical bosons part eventually turns out to be expressed through 
the single function $G(q_0)$. Up to the overall coupling 
constant, it reads \cite{IS},
\be
G(q_0) \propto \int du dv \frac{1 - X}{(1+X)^3} = 
2 (q^{ia}_0 q_{0\;ia})^{-1}.
\ee 
The component action, in the appropriate parametrization, coincides 
with the standard $(4,4)$, $SU(2)\times U(1)$ WZNW action.

It is worth mentioning that the action \p{confact} uniquely 
follows from requiring invariance under one of two different 
(though isomorphic) $N=4$ $SU(2)$ superconformal groups which one may 
realize in the $SU(2)\times SU(2)$ analytic HSS \cite{{DS},{IS}} 
(they close on the ``large'' $N=4$ 
$SO(4)\times U(1)$ superconformal group \cite{{belg1},{IKL}}). 
Also notice that there exist a few equivalent forms of the superfield 
lagrangian in \p{confact} which differ from each other by full 
harmonic derivatives. As an example we give two such forms of the 
$SU(2)\times U(1)$ WZNW action 
\bea
S_{wznw} = - {1\over \kappa^{2}} \int \mu^{-2,-2} 
\left(\hat{q}^{1,1} \right)^2 
\frac{c^{-1,1}c^{1,-1}}{(1+X)^2}  
= - {1\over \kappa^2} \int \mu^{-2,-2}\; 
c^{1,1}\hat{q}^{1,1} \frac{1}{1+X} \;. 
\label{confact1}
\eea  

Let us discuss massive deformations of the action \p{qact}. 
Surprisingly, and this 
is a crucial difference of the considered $(4,4)$ case from, say, 
$(2,2)$ sigma models with torsion, the only massive term of 
$q^{1,1\;M}$ 
consistent with analyticity and off-shell $(4,4)$ supersymmetry (not 
modified by central charges) is the following one \cite{{GI},{IS}}
\be 
S_m = m \int \mu^{-2,-2}\; 
\theta^{1,0\;\underline{i}} 
\theta^{0,1\;\underline{b}}\; 
C_{\underline{i}\;\underline{b}}^M\; 
q^{1,1\;M}
\;; \;\; [m] = cm^{-1}\;.
\label{mtermg}
\ee
Here $C_{\underline{i}\;\underline{b}}^M $ are arbitrary 
constants (subject to the appropriate reality conditions). 

After adding the term \p{mtermg} to the action \p{qact}, passing to 
components and eliminating auxiliary fields, the effective addition 
to the $(4,4)$ sigma model component action is given by 
\be
S_{q}^{pot} = {{m^2}\over 2} \int d^2 z \;G^{M\;N}(q_0)\;
(C_{\underline{i}\;\underline{a}}^M C^{\;\underline{i}\;
\underline{a}\;N}) \;.
\ee 
Here $G^{M\;N}(q_0)$ is the inverse of $G_{M\;N}(q_0)$ defined  
in eq. \p{auxfun}. Thus we see that the potential term in the case in 
question is uniquely determined by the 
form of the bosonic target metric. In particular, in the case of 
$(4,4)$ $SU(2)\times U(1)$ WZNW model one gets the Liouville 
potential term for 
$u(x)\propto \mbox{ln} \{ \mbox{tr}\; q^{ia}_0 \}$, so the 
massive deformation of this model is nothing but 
the $(4,4)$ super Liouville theory \cite{IK}. It would be 
interesting to 
inquire whether $(4,4)$ extensions of other integrable $2D$ theories 
can be obtained as massive deformations of some appropriate  
$(4,4)$ sigma models with torsion. 

\vspace{0.3cm}
\noindent{\bf 3. A dual formulation of the twisted multiplet 
and its generalization.} The above $SU(2)\times SU(2)$ HSS 
description of 
$(4,4)$ twisted multiplet suggests a new off-shell formulation  
of the latter via unconstrained analytic superfields. After 
implementing  
the constraints \p{qconstr} in the action with superfield lagrange 
multipliers and adding this term to \p{qact} we arrive at 
the following new action \cite{IS}
\be
S_{q,\omega} = \int \mu^{-2,-2} \{ 
q^{1,1\;M}(\;D^{2,0} \omega^{-1,1\;M}   + 
D^{0,2}\omega^{1,-1\;M} \;) + h^{2,2} (q^{1,1}, u, v) \}\;.
\label{dualq}
\ee
In (\ref{dualq}) all the involved superfields are unconstrained 
analytic, so from the beginning the action (\ref{dualq}) 
contains an infinite number 
of auxiliary fields coming from the double harmonic expansions with 
respect to the harmonics $u^{\pm1\;i}, v^{\pm1\;a}$. Varying 
with respect to the superfields $\omega^{1,-1\;M}, 
\omega^{-1,1\;M}$ takes one back to the action \p{qact} and 
constraints \p{qconstr}. On the other hand, varying with respect to 
$q^{1,1\;M}$ yields an algebraic equation for eliminating this 
superfield. This enables one to get a new dual off-shell 
representation of the twisted 
multiplet action through unconstrained analytic superfields 
$\omega^{-1,1\;M}$, $\omega^{1,-1\;M}$. 

The crucial feature of the action (\ref{dualq}) (and its $\omega$ 
representation) is the abelian gauge invariance 
\be  
\delta \;\omega^{1,-1\;M} = D^{2,0} \sigma^{-1,-1\;M} 
\;, \; \delta \;\omega^{-1,1\;M}  
= - D^{0,2} \sigma^{-1,-1\;M}\;,
\label{gauge}
\ee
where $\sigma^{-1,-1\;M}$ are unconstrained analytic 
superfield parameters. This gauge freedom ensures the 
on-shell equivalence of the $q,\omega$ or $\omega$ formulations of 
the twisted multiplet action to its original $q$ formulation 
\p{qact}. Namely, it 
neutralizes superfluous physical dimension component fields in the 
superfields $\omega^{1,-1\;M}$ and $\omega^{-1,1\;M}$ and 
thus equalizes the number of propagating fields in both 
formulations. It 
holds already at the free level, with 
$h^{2,2}$ quadratic in $q^{1,1\;M}$. So it is natural to expect that 
any reasonable generalization of the action (\ref{dualq}) respects 
this symmetry or a generalization of it. 
  
It is well known that with making use of $(4,4)$ twisted multiplets 
one may construct invariant off-shell actions only for those 
torsionful $(4,4)$ 
sigma models for which left and right triplets of covariantly constant 
complex structures on the bosonic target mutually commute 
\cite{{GHR},{RSS}}. 
This is true of course for both the actions \p{qact} and \p{dualq}. 
However, it turns out that the second one is 
a good starting point for constructing more general actions. They 
admit no inverse duality transformation to the twisted multiplets 
actions 
and yield an off-shell description of sigma models with non-commuting 
left and right complex structures. 

In refs. \cite{Iv} we started from the most general analytic superspace 
action of the triple of superfields 
$q^{1,1\;M}, \omega^{1,-1\;M}, \omega^{-1,1\;M}$. Exploiting 
the freedom of target space reparametrizations together with 
the constraints 
which stem from the important self-consistency condition 
\be \label{comm} 
[\;D^{2,0}, D^{0,2}\;] = 0\;,
\ee
we reduced the action to the form 
\bea
S_{q,\omega} &=& 
\int \mu^{-2,-2} \{\; q^{1,1\;M}D^{0,2}\omega^{1,-1\;M} + 
q^{1,1\;M}D^{2,0}\omega^{-1,1\;M} +  \omega^{1,-1\;M}h^{1,3\;M} 
\nonumber \\
&&+ \omega^{-1,1\;M}h^{3,1\;M} + \omega^{-1,1\;M} \omega^{1,-1\;N}
\;h^{2,2\;[M,N]} + h^{2,2}\;\}\;. \label{haction}
\eea
Here, the involved potentials depend only on $q^{1,1\;M}$ and target 
harmonics and still satisfy some additional constraints following from 
eq. \p{comm}. Most important of them is as follows 
\bea 
h^{2,2\;[N,T]}\;\frac{\partial h^{2,2\;[M,L]}}{\partial q^{1,1\;T}} + 
h^{2,2\;[L,T]}\;\frac{\partial h^{2,2\;[N,M]}}{\partial q^{1,1\;T}} +
h^{2,2\;[M,T]}\;\frac{\partial h^{2,2\;[L,N]}}{\partial q^{1,1\;T}} 
\;=\; 0\;. 
\label{4} 
\eea
It ensures the action to be invariant under some 
{\it non-abelian} and {\it nonlinear} 
gauge transformations which generalize \p{gauge}. Their role is 
to maintain the correct number of physical fields ($4n$ bosons and 
$8n$ fermions). 
They affect not only the $\omega$ superfields, but $q^{1,1\;M}$ as well 
\bea
\delta q^{1,1\;M} = \sigma^{-1,-1\;N} h^{2,2\;[N,M]}\;. 
\label{gaugenab}
\eea
In general, these gauge 
transformations close with a field-dependent Lie bracket parameter: 
\be
\delta_{br} q^{1,1\;M} = \sigma^{-1,-1\;N}_{br} h^{2,2\;[N,M]}\;, \;\;
\sigma^{-1,-1\;N}_{br} = -\sigma^{-1,-1\;L}_1 \sigma^{-1,-1\;T}_2 
\frac{\partial h^{2,2\;[L,T]}}{\partial q^{1,1\;N}}\;.
\ee
We see that eq. (\ref{4}) guarantees the nonlinear closure of the 
algebra of gauge transformations (\ref{gaugenab}), and so it is a group 
condition similar to the Jacobi identity. Note that the non-abelian 
character of these transformations is directly related to the presence 
of the new non-vanishing potential $h^{2,2\;[N,M]}$.

It is a matter of straightforward computation to demonstrate that 
in the latter case the left and right 
complex structures on the bosonic target {\it do not commute}. We 
checked 
this property explicitly \cite{Iv} for a particular class of 
the above models corresponding to the ansatz 
\bea
h^{1,3\;N} = h^{3,1\;N} = 0\;, \; 
h^{2,2\;[N,M]} = b^{1,1} f^{NML} q^{1,1\;L}\;, \; 
b^{1,1} = 
b^{ia}u^1_iv^1_a, \; b^{ia} = \mbox{const}\;.\label{solut}
\eea
Here, the totally antisymmetric real constants $f^{NML}$ are 
structure constants of some $n$-dimensional semi-simple Lie algebra. 

Thus it is the presence of the potential $h^{2,2\;[N,M]}$ that leads to 
the above gauge symmetry and simultaneously to 
the non-commutativity of the left and right complex 
structures \cite{Iv}. So, {\it only for} $n\geq 2$ 
a new class of torsionful $(4,4)$ sigma models emerges. The action 
with non-vanishing $h^{2,2\;[N,M]}$ {\it does not} admit any 
duality transformation to the pure $q^{1,1\;M}$ form and involves 
an infinite number of auxiliary fields. 

Important problems ahead are to find out possible stringy 
applications of this new class of off-shell $(4,4)$ sigma models and 
to construct more general $(4,4)$ sigma model actions 
by incorporating other types of twisted $(4,4)$ multiplets within 
the $SU(2)\times SU(2)$ HSS (or its further extensions). It would 
be also interesting to couple these sigma models to $(4,4)$ 
supergravity.

\vspace{0.3cm}
\noindent{\bf 4. N=4 super KdV hierarchies.} A powerful way to construct 
generalized KdV hierarchies is to associate 
them to the proper conformal algebras and superalgebras as the second 
hamiltonian structure \cite{Gerv}. 

The second hamiltonian structure for $N=4$ superextensions of KdV is 
provided by $N=4$ SCA's. There exist two $N=4$ SCA's 
which are different in their affine Kac-Moody subalgebras: the minimal 
one with the $SU(2)$ affine subalgebra \cite{Adem} and a more 
extensive SCA with the subalgebra $SO(4)\times U(1)$ \cite{belg1}. 
Both of them and the associated super KdV hierarchies admit a natural 
formulation in the framework  of $N=4$, $1D$ HSS. 

The $N=4$, $SU(2)$ SCA is represented by the analytic harmonic 
spin $1$ supercurrent $V^{++}(\zeta)$, 
$\zeta = (z, \theta^{+ \underline{i}}, u^{\pm}_i)$, subjected 
to the harmonic constraint 
\be
D^{++} V^{++} =0\;,\;\;\;\; D^{++} = \partial^{++} + i
\theta^{+ \underline{i}} \theta^{+}_{\underline{i}} \partial_z 
\label{constrV}
\ee
(the notation is basically the same as in the previous sections). 
With this constraint, the superfield $V^{++}$ displays the irreducible 
current contents of $N=4$, $SU(2)$ SCA: the spin $1$ affine current
$v^{(ik)}$, the spin $3/2$ fermionic current $\xi^{i\underline{k}}$ and 
the spin $2$ conformal stress-tensor $T$. It is easy to write superfield 
Poisson brackets for $V^{++}$ which reproduce $N=4$, $SU(2)$ SCA for 
these currents \cite{DI}
\bea  \left\{ V^{++}(1),
 V^{++}(2)\right\} &=& {\cal D}^{(++|++)} \Delta (1 - 2) \nn \\
{\cal D}^{(++|++)} &\equiv & (D^+_1)^2(D^+_2)^2 \left(
\left[ \left({u^+_1u^-_2
\over u^+_1u^+_2}\right) - {1\over 2}D^{--}_2\right]
 V^{++}(2)
- {k\over 4}\pl_{2}\right).
\label{poi}\eea
Here $\Delta(1-2)=\delta(x_1-x_2)\;(\theta^1-\theta^2)^4$ is the
ordinary $1D \;N=4$ superspace delta function and 
$$
(D^+)^2 \equiv D^+ \bar D^+\;.
$$
Note that the harmonic
singularity in the r.h.s. of \p{poi} is fake: it is cancelled after
decomposing the harmonics $u^{\pm i}_2$ over $u^{\pm i}_1$ with making
use of the completeness relation 
$u^+_iu^-_k - u^+_ku^-_i = \epsilon_{ik}$. 

Now it is straightforward to derive 
the relevant evolution equation, the $N=4$, $SU(2)$ super KdV equation 
\cite{DI} 
\be  \partial_t V^{++} = \left\{ H_3, V^{++} \right\}\;,\ee
\be
H_3 =\int [dZ]\;V^{++}(D^{--})^2V^{++}-i\int [d\zeta^{-2}]\;
c^{-4}(u)\;(V^{++})^3
\;.
\label{h3}
\ee
Here, $[dZ]$ and $[d\zeta^{-2}]$ are appropriate integration measures 
over the full $1D$ HSS and its analytic subspace, $D^{--}$ is the 
second harmonic 
derivative (it does not preserve the analyticity) and $c^{-4} = 
c^{(ijkl)}u^{-}_i u^{-}_j u^{-}_k u^{-}_l$ is a $SU(2)$ breaking 
tensor (in the explicit form, this equation is given in 
\cite{{DI},{DIK}}). The $N=4$ super KdV is integrable (generates the 
whole hierarchy and is bi-hamiltonian) under the 
following restrictions on $c^{(ijkl)}$ \cite{{DI},{DIK}}
\be
(a)\;c^{(ijkl)} = {1\over 3} \left( a^{(ij)} a^{(kl)} + 
a^{(ik)}a^{(jl)} + 
a^{(il)}a^{(jk)} 
\right); \;\; (b)\;a^{(ij)}a_{(ij)} \propto
-{1\over k}\;,
\label{ct4} \ee
$k$ being the level of $SU(2)$ Kac-Moody subalgebra. There exist two 
non-equivalent reductions of this system to $N=2$ super KdV, yielding 
the $a=4$ and $a=-2$ integrable cases \cite{Mat1} of the latter. 
An interesting unsolved problem is to construct the Lax representation 
for this $N=4$ super KdV.

The $N=4$, $SO(4)\times U(1)$ SCA is also tightly related to 
$1D$ harmonic analyticity, this time to the $SU(2)\times SU(2)$ one 
\cite{DI1}. 
Its basic object is a spin $1/2$ fermionic supercurrent $J^{1,1}$. It 
lives on the three-theta analytic subspace of the 
$SU(2)\times SU(2)$, $1D$ HSS
\be
D^{1,1}J^{1,1} = 0 \Leftrightarrow J^{1,1} = J^{1,1}(\xi_3)\;, 
\;\;\xi_3 = 
(z, \theta^{1,1}, \theta^{1,-1}, \theta^{-1,1}, u^{\pm 1}_i, 
v^{\pm 1}_{\underline{k}} )\;,    \label{3anal}
\ee
(with $D^{\pm1, \pm1} 
\equiv D^{i\underline{k}}u^{\pm1}_iv^{\pm1}_{\underline{k}}$ 
in the central basis) and obeys the harmonic constraints 
\be
D^{2,0}J^{1,1} = D^{0,2} J^{1,1} = 0\;. \label{qconstr2}
\ee 
Its irreducible field contents include the spin $1/2$ current 
$j^{i\underline{j}}$, the affine $SO(4)\times U(1)$ spin $1$ currents 
$v^{(ik)}, v^{(\underline{i}\underline{j})}, v$, the spin $3/2$ 
currents
$\xi^{i\underline{k}}$ and the stress-tensor $T$. This is just the set 
forming $N=4$, $SO(4)\times U(1)$ SCA. It is easy to establish the 
proper Poisson brackets between $J^{1,1}$'s and to write the related 
superfield evolution equation 
\be
\partial_t J^{1,1} = \left\{ H^{'}_3, J^{1,1} \right\}\;, \label{lN4kdv}
\ee
$H^{'}_3$ being the most general dimension $3$ $N=4$ supersymmetric 
hamiltonian composed of $J^{1,1}$ and its derivatives. This equation 
is a 
kind of the ``master'' one, as all the previously known super KdV 
equations are expected to follow from it via proper reductions. 
For instance, 
the supercurrents of two $N=4$, $SU(2)$ SCA's present in 
$N=4$, $SO(4)\times U(1)$ SCA are defined as 
\be 
V^{2,0} \equiv D^{1,-1}J^{1,1},\; \; V^{0,2} \equiv D^{-1,1}J^{1,1}\;.
\ee
Besides the analyticity condition \p{3anal}, they meet extra 
analyticities 
\be 
D^{1,-1}V^{2,0}= 0,\;\; D^{-1,1}V^{0,2} = 0, 
\ee
which imply them to live on two different two-theta analytic 
subspaces of the $SU(2)\times SU(2)$, $1D$ HSS
\bea
V^{2,0} &=& V^{2,0} (\xi_2),\; \;V^{2,0} \;=\; V^{2,0} (\xi_2^{'}), 
\nonumber \\
\xi_2 &=& (z^{'}, \theta^{1,1}, \theta^{1,-1}, u^{\pm 1}_i,
v^{\pm 1}_{\underline{k}}), \; \;
\xi_2^{'} \;=\; (z^{''}, \theta^{1,1}, \theta^{-1,1}, 
u^{\pm 1}_i, v^{\pm 1}_{\underline{k}})\;.
\eea
Both supercurrents satisfy the bi-harmonic shortness conditions 
like \p{qconstr2}. In the limits $V^{2,0} = 0$ or $V^{0,2} = 0$ the 
remaining 
supercurrent is expected to satisfy the $N=4$, $SU(2)$ KdV equation as 
a consequence of the $SO(4)\times U(1)$ one \p{lN4kdv}. The analysis 
of self-consistency of these reductions as well as the 
issue of integrability of \p{lN4kdv} (the existence of an infinite 
set of conserved quantities, Lax representation, ...) 
are now under study.

\vspace{0.3cm}
\noindent{\bf Acknowledgment.} I thank Organizers of the Alushta 
Conference for offering me an opportunity to give this Talk. This work 
was supported by RFFI grant RFFI 96-02-17634, INTAS grant INTAS-94-2317 
and by a grant of the Dutch NWO organization.

\end{document}